# University-industry collaboration in Italy: a bibliometric examination[*][1]

**Abstract**

This work investigates public-private research collaboration between Italian universities and domestic industry, applying a bibliometric type of approach.

The study is based on an exhaustive listing of all co-authored publications in international journals that are jointly realized by Italian university scientists and researchers in the private sector: this listing permits the development of a national mapping system for public-private collaboration, which results unique for its extensive and representative character. It is shown that, in absolute terms, most collaborations occur in medicine and chemistry, while it is industrial and information engineering which shows the highest percentage of co-authored articles out of all articles in the field.

In addition, the investigation empirically examines and tests several hypotheses concerning the qualitative-quantitative impact of collaboration on the scientific production of individual university researchers. The analyses demonstrate that university researchers who collaborate with those in the private sector show research performance that is superior to that of colleagues who are not involved in such collaboration. But the impact factor of journals publishing academic articles co-authored by industry is generally lower than that concerning co-authorships with other entities. Finally, a further specific elaboration also reveals that publications with public-private co-authorship do not show a level of multidisciplinarity that is significantly different than that of other publications.

[*] Authors are grateful to two anonymous referees for their comments and suggestions.
[1] Abramo, G., D'Angelo, C.A., Di Costa, F., Solazzi, M. (2009). University-industry collaboration in Italy: a bibliometric examination. *Technovation*, 29(6-7), 498-507. DOI: 10.1016/j.technovation.2008.11.003

# 1. Introduction

The capacity of a nation to produce wealth depends increasingly on the investment it undertakes in strengthening the so-called "triangle of knowledge", which is composed of research, education and innovation. In this regard, European nations, in accepting the Lisbon 2000 agenda, assumed an ambitious objective: to make Europe the most competitive and dynamic knowledge-based economic system in the world. The strategy, as further consolidated in the Barcelona summit, set the objective of EU member states assigning 3% of GDP to research by the year 2010. These directives indicate the desire to remedy Europe's competitive weaknesses at the international level.

As a vehicle for action, both the attention of policy makers and the accompanying debate seem heavily focused on the research apex of the research-innovation-development triad, with the Barcelona summit objectives (though recognized as a difficult target for most EU nations) clearly emphasizing the provision of resources for research. And yet the existence of a "European paradox" is well known, meaning that there is an incapacity to translate the excellent results from European research into innovations that are successfully destined for the marketplace (EC, 1995). Many analyses, comparing to the reference experience of North America, show that the greater competitive capacity of the nations there has clearly been favored by policies and legislation (such as the Bayh-Dole Act in the United States) which have stimulated technological transfer and provided incentive for osmosis between the worlds of public research and industry (Shane 2004; Thursby and Thursby, 2003; Mowery et al., 2001).

Within Europe then, the difference between the levels of scientific performance and technological and industrial competitiveness is most pronounced in nations such as Italy, where the government's expenditure in research is higher (50.7%) than the private sector's, industry is primarily specialized in low and middle-low technology, and the industrial structure composes a disproportionate number of micro and small enterprises. In the Italian context it is even more urgent that the nation promote collaboration between the public research system and industry, thus creating favorable conditions for commercial exploitation of the research results from universities and public research laboratories (Grandi and Sobrero, 2005). But, in observation, Italy registers a low propensity to capitalize on the results of public research. In 2001 (after that date, the introduction of academic privilege in Italian patent legislation would make international comparisons uneven), for example, the number of patents by the entirety of Italian universities was roughly equal to that of the University of Wisconsin alone, while that of the totality of all universities plus all public research laboratories was inferior to that of the Massachusetts Institute of Technology (Abramo and Pugini, 2005). As context for the comparisons, it may be worth knowing that the R&D expenditures in 2001 were around 486 million euro at MIT; 675 million euro at the University of Wisconsin; and 4,418 million euro at Italian universities (NSF, 2003; Istat, 2002).

To this must be added the limited capacity for transfer of patents to the productive system: the National Research Council, the major Italian research institution, awards licenses for the actual use of less than 20% of the patents it files annually; Italian universities award licenses for an average of 13% of patents, compared to 60% for universities in the United States and the UK (Abramo, 2007). Yet a 2005 study by Abramo and D'Angelo that examined the alignment of public research supply with Italian industrial demand, through a survey of leading public research scientists in high-tech sectors, found that most research project results do seem to have immediate industrial applicability, even if in one third of the cases there are no Italian companies able to exploit the results. The ensemble of these observations points to the clear necessity of fine-tuning the match between research policy and industrial policy, with greater attention to all initiatives that may foster the transfer of public research results to domestic industry.

   The relations between universities and industry presently take form in various modes, variable in the extent to which they are codified and formalized. Typical modalities include joint research projects, awarding of research contracts, awarding of know-how and patents under license, consulting, training services and personnel mobility. The observation of such modalities, their empirical study and the analysis of their underlying determinants can furnish useful cognitive bases for the policy maker called to stimulate them. In this regard, the present study proposes to investigate research collaboration between universities and domestic industry through a bibliometric approach, in which "collaboration" is represented by "co-authorship of scientific articles", and to develop a mapping system able to identify the technical-scientific fields in which alignment between private demand and public offer of knowledge is realized with greater frequency. In a complementary manner, such mapping will highlight those sectors in which the connection between academic and productive systems is weak or completely absent. This information may result as useful for the policy maker, both for choosing the development directions and aims for programming sectorial priorities, as well as in monitoring results from previous interventions with similar objectives (Owen-Smith et al., 2002; Veugelers and Cassiman, 2005).

   Another relevant aspect of the proposed investigations concerns the analysis of the determinants of collaboration for public-private research: these can be considered as exchange relationship in which both parts obtain benefits (Meyer-Krahmer and Schmoch, 1998). In particular, on the part of university researchers, collaboration with private business guarantees access to additional financing for research and/or to complementary assets. It should be noted that the correlation between defined objectives *ex ante* and benefits obtained *ex post* is not always linear and that many benefits are obtained in an unexpected manner (Lee, 2000). All this should have a significant impact on qualitative-quantitative productivity of scientists (Balconi and Laboranti, 2006; Barnes et al., 2002; Van Looy et al., 2004). However, since collaboration involves interaction between individuals, and in the case of public-private cooperation, between individuals appertaining to systems that are very different in their identity and mission, it brings about transaction costs. These are costs resulting from needs to negotiate and mediate objectives, choose methodologies, deal with results, manage logistics for communications, manage gatherings and face-to-face meetings, and for further coordination needs, and they are costs that would logically create disincentives towards collaboration (Belkhodja and Landry, 2005; Drejer and Jorgensen, 2005). In effect, a vast survey conducted in Great Britain by D'Este and Patel (2007) showed that the determinants of the variety and frequency of public-private interactions depend above all on the individual characteristics of the researchers involved, more so than the characteristics of their home organizations. Most importantly, it seems there is little evidence of conflict between interactions with industry and more traditional academic roles (Boardman and Ponomariov, 2008).

In view of these potential benefits and transaction costs, another objective of this study is to test whether collaboration with the private sector actually produces scientific results that are qualitatively better (from the viewpoint of publication placement) and if academic scientists who collaborate with those in the private sector demonstrate superior performance with respect to colleagues who are not involved in such collaboration.

The last aspect for exploration concerns the multidisciplinarity of projects undertaken in cooperation between academic scientists and private sector researchers. It can be hypothesized that projects that interest private companies necessarily imply, by their nature, a call for heterogeneous and varied competencies. This weighs on transaction costs, which will clearly be variable with the level of heterogeneity among the members of a mixed research team. Thus, the last objective to be pursued in this study is to examine and verify whether there is a higher level of multidisciplinarity in public-private research projects, which could determine a surplus transaction cost for this type of project and a consequent disincentive for the actors in play, in particular for academic scientists.

From a methodological perspective, this study, although limited to observation of the Italian situation, is characterized by its very ample field of analysis, both for the number of academic institutions (all 68 Italian universities) and for the scientific sectors analyzed (the full 183 sectors of the 8 technical-scientific areas of the Italian academic system). This constitutes an innovative aspect with respect to preceding studies, which have generally been based on partial measures of one or a few universities, and/or have focused on single scientific sectors.

The authors are aware that co-authorship based indicators should be handled with care as a source of evidence for true scientific collaboration, as has been cautioned by many bibliometricians (Melin and Persson, 1997; Laudel, 2002; Tijssen, 2004; Lundgerg et al., 2006). As Katz and Martin (1997) stated, some forms of collaboration do not generate co-authored articles (university researchers might for example publish without mentioning the direct involvement of industrial researchers) and some co-authored articles do not reflect actual collaboration (a publication could suggest an inter-institutional collaboration that has not taken place, for example if an author has moved from a university to industry and in his/her publication lists both the prior and current affiliation).

However it is incontestable that, in the literature, analysis of co-authorship has become one of the standard ways of measuring research collaborations between organizations, evidently because it offers notable advantages in counterpoint to the limitations noted above. Co-authored publications indicate the achievement of access to an often informal network, and can be viewed as successful scientific collaboration in themselves, while also indicating diffusion of knowledge and skills. Moreover the indicator is quantifiable and invariant, while measurement is not invasive and analysis is relatively inexpensive. Finally, with reference to the specific character of the study proposed, the numerous cases observable as proxy (more than 1500 publications, for a total of almost 2000 collaborations in the 2000-2003 triennium under examination) certainly guarantee a level of significance that could not be reached through alternative approaches, for example those based on listings of patents authored by academic scientists but owned by private firms, or on sample-based surveys.

The next section of this report presents the data set used in the study, while Section 3 depicts the mapping of collaboration, by area and disciplinary sector. Section 4, in reference to the second objective of the study, presents the analysis of the qualitative-quantitative impact of collaboration with private sector colleagues on the research performance of academic scientists. Section 5 explores the level of multidisciplinarity of research projects in private-public co-authorship, while the last section closes the work with a brief synthesis and the final thoughts of the authors.

## 2. *Dataset*

As noted in the introduction, the investigation of the phenomenon of collaboration in academic research typically considers scientific publications in international journals that are co-authored by universities with any other type of organization: other universities, public research laboratories, domestic companies, organizations from other nations, etc. However, for the objectives of this particular study, the data set under specific investigation consists of publications in co-authorship with domestic industry.

The source of reference is the Observatory of Public Research (*Osservatorio sulla Ricerca Pubblica,* or ORP) which registers, for the 2001 to 2003 triennium, the international scientific production of all Italian universities. The ORP is in turn based on the data of the Thomson Scientific SCI™, Cd-Rom version. In order to assess to what extent the SCI$^{TM}$ and consequently ORP can serve as representative of the academic research outputs in the "hard" sciences, a verification was made by Abramo et al. (2008). The articles in international journals indexed in ORP amount to an average of 95% of the total outputs submitted by the Italian universities in the first and only Italian research evaluation exercise. Selecting every listed publication with at least one address corresponding to an Italian university, the ORP then applies a disambiguation algorithm to attribute the publication to its respective academic authors. For details see Abramo et al., 2008. Since Italian university research personnel are subdivided by scientific disciplinary sector (SDS), it is possible to link each publication (and each collaboration) to the SDSs to which the university authors appertain. The SDSs are grouped in macro university disciplinary areas (UDAs). The field of observation for the present analysis considers 8 technical-scientific UDAs (mathematics and computer sciences, physics, chemistry, earth sciences, biology, medicine, agricultural and veterinary sciences, industrial and information engineering) including 183 SDSs. This level of detail permits overcoming several distortions typical of aggregate analyses that do not give due consideration to the different "fertility" of scientific disciplines and the different representivity by discipline within the journals that are listed in the source database (Abramo et al., 2007).

The task of listing the publications of interest, i.e. those co-authored by universities and domestic companies, also imposed the identification and consistent rendition of all the possible names of domestic firms present in the address field of publications listed in the ORP.

The work here is unique with respect to the international state of the art for at least two features, firstly for its broad field of observation: studies in the previous literature have only been based on limited samples of the population of interest, and have tended to focus on restricted disciplinary sectors or single institutions. Instead of these approaches, the study proposed here refers to the entire population of all academic research scientists from all technological-scientific fields, being a total of 33,000 scientists. Secondly, the study is unique for the method used, of categorizing each collaboration and comparing individual performance: each scientist has been individually identified, then classified and grouped by role and scientific field of specialization. This permits the limitation of otherwise inevitable distortions in productivity measurement due to non-homogeneity of units under comparison (see Abramo and D'Angelo, 2007). The analysis is based on the entire population of Italian university research staff and thus avoids problems in robustness and significance of inferential analyses. It further presents an undeniable advantage of objectivity and homogeneity in the source data, not always found in examinations based on questionnaires.

### 3. Sectorial mapping of university-industry collaboration in research

Overall, there were 791 domestic companies (the legal entities considered are private companies located in Italian territory. The following have been excluded: publicly owned organizations, mixed public-private consortiums and foundations) in the 2001-2003 triennium which realized at least one international scientific publication listing in the ORP. Of these, 483 collaborated at least once with an Italian university. On the other side, 63 out of 68 universities collaborated with industry, in the areas under examination. Such collaboration, resulted in 1,534 articles, approximately 3% of the over 52,000 articles bearing the names of university researchers. Each article can indicate more than one collaboration, in function of the number of universities and private firms present in the address field of the article itself. There are 4 possible cases:
- 1 university, 1 corporation = 1 collaboration
- m universities, 1 corporation = m collaborations
- 1 university, n private firms = n collaborations
- m universities, n corporations = m x n collaborations

As a whole the 1,534 co-authored articles embed 1,983 collaborations, of which 1,195 (60%) are of the first type, 646 (33%) of the second type, 92 (5%) of the third type and 50 (2%) of the fourth type.

To quantify the level of intensity of collaboration between universities and private companies in the various scientific sectors, four types of indicators were taken into consideration:
- the number of university articles in co-authorship with private researchers, in a given SDS/UDA;
- the percentage of articles in co-authorship with private researchers, out of the total of articles realized in the specific SDS/UDA;
- the percentage of articles in co-authorship with private researchers, out of the total articles realized in co-authorship, in the specific SDS/UDA. By this indicator we can see to what extent public-private collaboration is sector specific.
- The number of articles in co-authorship with industry per researcher in the specific SDS.

Table 1 presents the data relative to the analysis by disciplinary area. Double counting of articles may occur here because an article may fall in more than one disciplinary area. In terms of mass (number of articles in co-authorship), the medicine and chemistry areas dominate. Referring to the other two (normalized) indicators, it is industrial and information engineering that leads, and the ranking of the first 4 disciplinary areas is invariant: for industrial and information engineering, over 6% of publications bear the joint signature of university scientists and researchers from private firms. In quite distant second place arrives chemistry (3.9%) and in sequence, agricultural and veterinary sciences and biology (2.8%). Medicine, which leads the rankings for number of articles in university-industry co-authorship places below fourth position in the normalized ranking for overall scientific production.

[Table 1]

The details by individual scientific disciplinary sector (SDS) are presented in Tables 2, 3 and 4. Ahead of all others, electronics is the sector with the most articles with co-authorship between universities and corporations, numbering a full 114 (Table 2), followed next by internal medicine (109), pharmacology (94) and biochemistry (89). Below fourth position we find 5 disciplinary sectors from the chemistry area and one from physics (experimental physics).

[Table 2]

Considering, in each SDS, the rating for incidence of articles in co-authorship with the private researchers as a percentage of total scientific production by all university scientists in the same SDS (Table 3), electronics recedes to third place (12.6%), overtaken by the sector of energy and environmental systems (15.0%) and by polymer materials science and technology (13.3%). In the first 10 positions, we find 8 sectors from the industrial and information engineering disciplinary area and two from chemistry.

The domination of the industrial and information engineering area is also observed in the rankings for percentage of articles realized in co-authorship with private sector researchers out of the total of articles with co-authorship (Table 4). In the first 10 positions, a full nine, and among these the first 7, are occupied by SDSs from this area. In general, it is possible to observe that the sectors concerned, as could be expected, are those directed towards applied science. This can probably be retraced to the structure of the Italian productive system, which primarily articulates around small and medium enterprises, operating for the most part in "non-high-tech" areas and therefore more inclined to collaborate with universities if this involves research projects with a prevalently practical application.

Relating the scientific production realized in co-authorship with private sector researchers to the number of university researchers on staff in each SDS brings out the data seen in Table 5. These depict a situation that is not different from that emerging from Table 3, with the first 10 positions including five SDSs from the chemistry area, four from industrial and information engineering and one, the last from biology (molecular biology). Also, allowing for the exception of the electronics discipline, the other three SDSs from the industrial and information engineering area are linked with the field of chemistry (particularly applied chemistry), making this the area with the highest intensity of scientific collaboration with private firms, per single university researcher.

[Table 3]
[Table 4]
[Table 5]

## 4. University-industry collaboration and quality of output

### 4.1 Impact of university-industry co-authored publications

The university-industry collaboration that is the object of this study brings together individuals from two distant worlds: public research institutes and private industry are characterized by highly divergent missions, organizational structures and management systems. The term "ivory tower" (Zuckerman, 1971) gives an image of the limited permeability of universities to the external world, including to demands for new knowledge that might arise in industry. Therefore, for those public researchers willing to take on the challenge of realizing collaboration with the private sector, such collaboration must present significant strategic, economic or financial returns. The public researcher's election to collaborate with the private sector must clearly be linked to personal interests and benefits: first to possibilities of obtaining financing, access to physical assets and complementary competencies, leading to the possibility of achieving significant results that would further add to personal visibility and prestige. This brings about formulation of a hypothesis concerning the quality of results that can be obtained by the university researcher: is it possible that results on average are significantly superior when originating from collaboration with industry?

The intention of this part of the study is thus to test for the existence of a potential differential in average quality between the totality of research products realized by universities in the period under observation and that obtained specifically as a result of collaboration with private firms. For this, the study refers to the publication placement of such products in the scientific journals of the various sectors, expressed as the Impact Factor (IF) of the relevant journals. This indicator represents a proxy measure of the "quality" of the journal rather than that of the article itself. The authors are aware of the intrinsic limitations of such approximation, as well as of the recommendations contained in the literature on this issue (Moed and Van Leeuwen, 1996, Weingart, 2004). However, as they did not have access to data on the actual number of citations of each article and as their purpose is to compare two subsets of the same population the authors decided to proceed with the comparison on the basis of the proxy measures. To give due consideration to sectorial variability, in terms of number of journals and of the distribution of IF for all the journals of each SCI scientific category, the value of IF for each journal will first be transformed into its percentile rank ($IF_{pr}$) within the distribution of IFs of the journals in the same scientific category.

The results of the analysis are illustrated in Tables 6 and 7. In particular, Table 6 first presents the results of comparison between the universe and only the publications realized in collaboration: the data clearly indicate that the publication placement and the impact of the publications realized in collaboration are significantly superior compared to that of the totality of publications listed in the period under observation. Further, in 148 SDSs (of the 181 that illustrate at least one collaboration between a university and any other type of organization), the publications from such collaboration show an impact index superior to that resulting from the totality of the publications.

[Table 6]

Next, the comparison conducted between all the publications and those realized only in collaboration with private sector researchers (Table 7): this comparison instead shows that the difference between the average percentile ranks of IF for these two data sets is not statistically significant. In the 142 SDSs which register at least one collaboration with the private sector, only 65 (circa 46%) of the publications with private sector researchers show an impact superior to that for all the publications.

[Table 7]

Thus, if co-authorship or collaboration in general seems to relate positively to the publication placement of the results that can be obtained from university research activity, this does not seem to hold true when the collaboration by university researchers is specifically with those colleagues who appertain to private firms.

## 4.2 Collaboration and performance of university researchers

The preceding section showed that the difference in quality of research results obtained by university scientists should not constitute a significant incentive for collaboration with industry. Scientific production deriving from university-industry collaboration does not achieve positioning in particularly prestigious journals.

It is necessary to verify the existence (or lack of) another potential incentive for collaboration between private and university researchers: access to additional financing, physical assets and complementary know-how could determine (all else being equal) a significant increment in quantitative productivity for the university researchers. It is clear that collaboration entails (in Italy at least) a task of re-adaptation by university researchers due to the necessity of sharing objectives, programs and operational lines with their private partners. Such adaptation, if on the one hand repaid in terms of financial support, could on the other hand create an element of "distraction" with respect to the institutional duties of the scientist. To examine which of the two incentives prevails, either the centripetal (which favors collaboration) or the centrifugal (which contrives against), this section will compare and contrast the performance of individual university researchers who partner with private sector colleagues, compared to the rest of their colleagues in the sector.

An alternative explanation to interpret any observation of higher productivity by authors involved in public-private cooperation could be that of efficient selection on the part of private firms. In other words, the explanation would be that industry chooses to collaborate with the best university researchers.

For this examination, two indicators will be taken in consideration:
- Output ($O$): the sum of publications realized by the scientist in the triennium under

consideration;
- Fractional Scientific Strength (*FSS*): the weighted sum of contributions to publications realized by the scientist (the weight being equal to the normalized impact factor of the publishing journal and the contribution for each publication being considered as the inverse of the number of co-authors).

This second indicator takes into consideration all three of the relevant dimensions of individual performance: quantitative (number of publications), qualitative (impact of the publication journal) and contributive (number of co-authors).

When proceeding to comparisons with aggregated data, the performance registered by single scientists is provided in terms of percentile ranking in the distributions of their respective sectors ($O_{pr}$, $FSS_{pr}$). Of the 17,857 Italian university researchers on staff in those technical-scientific sectors that realized at least one scientific article (in the 2000-2003 triennium), 1705 researchers (approximately 10%) published in co-authorship with private sector scientists (Table 8).

[Table 8]

This set of scientists demonstrates performance significantly superior to that of their remaining colleagues: in terms of output, the gap in performance is less than 26.8%; for Fractional Scientific Strength the gap rises to 29.4%.

This comparison thus demonstrates that, without question, university researchers involved in research partnerships with industry have qualitative-quantitative scientific performances that are invariably higher than those of the rest of their colleagues in the same sector.

## 4.3 Collaboration and multidisciplinarity

The study of level of multidisciplinarity in public-private collaboration offers interesting points for reflection on the motivations that would stimulate a private corporation to call on a university. A search for collaboration could actually be dictated by the need to avail of a spectrum of competencies in sectors far removed from one another, to favor or accomplish cross-fertilization phenomena which may foster innovation, and also to meet the demand for expertise in highly circumscribed and specialized fields (the "phenomenon of specialization"). From the perspective of the university researcher, the heterogeneity of competencies in any research team under consideration represents a potential opportunity, but also a determinant of an additional cost component with respect to the normal transaction costs typical of collaboration with second parties.

The study of the level of multidisciplinarity of joint public-private research projects thus also constitutes a useful basis to test for the eventual presence (or lack) of possible disincentives for university researchers to undertake scientific collaboration with private partners.

For the purpose, this study takes into consideration two different indicators of multidisciplinarity: one relative to the disciplinary sectors to which the university scientists adhere ($Ii\_{SDS}$), the other referring to the SCI scientific category associated with the journals listed in the SCI™ ($Ii\_{SCI}$).

$Ii\_{SDS}$ = average number of SDSs associated with the university researcher coauthors of the publications in the data set

$Ii\_{SCI}$ = average number of scientific categories associated with the SCI journals of the dataset

Table 9 presents statistics about SDS mean values of the above two indicators, respectively for all academic publications and for those in collaboration with private companies. Referring to the first indicator, it can be noted a significant difference between these two subsets of publications: 1.649 is the mean value of number of SDSs represented in all academic articles vs 1.763 in those in co-authorship with industry. Proceeding to a higher level of detail, it can be seen that in 75 sectors (of the 141 with at least one collaboration with industry) there is a higher index of multidisciplinarity $Ii\_{SDS}$ for publications in collaboration with industry researchers.

Referring to the other indicator $Ii\_{SCI}$, at the aggregate level the average number of disciplinary categories associated with the journals that publish articles with university-industry co-authorship is greater than that for the entire population (2,222 versus 2,111), although this difference seems not statistically significant. Classifying the publications according to the SCI category of the journal concerned and then proceeding to analyses by single SCI category, it emerges that in 79 cases (of the 144 with at least one publication in co-authorship between an Italian university and Italian corporation) the index of multidisciplinarity $Ii\_{SCI}$ is greater for publications in collaboration with private sector researchers; this does not hold true for the complementary 45% of cases.

[Table 9]

The same analysis was carried out for comparing the level of multidisciplinarity in publications in co-authorship with industry and all academic publications attained in extra-mural collaborations (Table 10). For both indicators the difference in SDS mean values in these two subsets is small and not statistically significant. In this case also, we recorded a certain variability in sectors data: the index of multidisciplinarity $Ii\__{SDS}$ is higher for publications in collaboration with industry in 66 SDSs (out of total 141); the index $Ii\__{SCI}$ is higher in 80 sectors (out of total 144).

[Table 10]

## 5. Conclusions

This study, by means of a bibliometric approach, has examined several salient features of research collaboration between the Italian universities and domestic industry.

The approach taken, although with the typical limitations of using co-authorship of scientific articles published in international journals as a proxy of collaboration, presents the advantage of being based on objective quantitative data and, above all, on a type of full census that permits exhaustive sectorial mapping of existent scientific cooperation between university researchers and their private sector colleagues. This mapping could serve as a first development of a useful aid for policy makers. In fact, regular input and systematic updating for this map would configure a true and unique observatory of university-industry collaboration in national research, useful for assessing, among others, the actual impact of relevant policies. Further, comparing the Italian situation on an international basis would be useful for identification of the strengths and weak points of the national system, with respect to the situation of benchmark nations. This is even more important in a nation such as Italy, where the government's expenditure in research is higher (50.7%) than the private sector's, industry is primarily specialized in low and middle-low technology, and the industrial structure is made of a disproportionate number of micro and small enterprises. In a nation with such given conditions, the best strategy to maintain pace with the leaders and avoid losing ground to emerging economies, in terms of technological competitiveness, would be that of providing incentive for the maximum exploitation of public research results by industry.

This study also permitted examination of several hypotheses about the relationship between university-industry collaboration and quality of research output. In particular, it emerged that university-industry articles do not have a better placement, in terms of the impact of the journal of publication, with respect to other publications. This thus seems to exclude the presence of potential incentive towards research partnership with industry aimed at improving the quality researcher output.

Still, it also emerged that university researchers that collaborate with the private sector had overall personal research performances (both higher output and fractional scientific strenght) superior to those of their colleagues that do not undertake cooperation. For an appropriate interpretation of this result it certainly remains to search for the direction of a causal relationship emerging from the data: is it the corporations that choose the best university researchers, or (given equal professional capacities) is the apparent higher research efficiency (albeit not in publications directly linked to the collaboration) for researchers who collaborate with the private sector actually due to financial support received and to the stimulus obtained from contact with a private partner? While this question will be the object of a specific supplementary investigation, Lee and Mansfeld (1996) in their survey of relations between a sample of high-profile American high-tech companies and universities, show that geographic distance results as an important factor in determining collaborations. Businesses tended to finance applied research in universities found within 100 miles of their base, even if these did not demonstrate high levels of excellence. Universities situated beyond this threshold had a greater possibility of being chosen only if they were among those with an optimal reputation. In such cases, the projects involved were usually "basic" research, in which proximity is less determining and scientific prestige of the faculty assumes greater importance. Because the collaborations that we analyzed are identified through the co-authorships of scientific articles, it is likely that companies tended to base their choice of the academic partners on scientific excellence.

Finally, the analyses conducted on the publications that are the fruit of university-industry collaboration reveal that these are characterized by a level of multidisciplinarity superior to the rest of the articles published by academic researchers. However no significant differences are observed when comparing all publications attained in extramural collaboration with those in co-authorship with private companies.

Further investigations could add to the results of this study by, for example, extending the analyses to other types of public research institutions and to the examination of cooperation with private firms located outside Italy. Finally, it would be interesting to carry out a more profound exploration of certain aspects of the geographic dimension of cooperation, referring particularly to factors influencing the choices made by corporations in selecting a public partner (such as the "proximity" effect), and to explore for trends of regional flow and spillover of knowledge generated by such collaboration.

|  | UDA 1 | UDA 2 | UDA 3 | UDA 4 |
|---|---|---|---|---|
| Number of articles in university-corporation co-authorship | Medicine (416) | Chemistry (415) | Industrial and inf. engineering (358) | Biology (308) |
| Percentage of articles in co-authorship with private sector out of the total UDA articles | Industrial and inf. engineering (6.4%) | Chemistry (3.9%) | Agricultural and veterinary sciences (2.8%) | Biology (2.8%) |
| Percentage of articles in co-authorship with private sector out of the total UDA articles with co-authorship | Industrial and inf. engineering (10.6%) | Chemistry (5.7%) | Agricultural and veterinary sciences (4.4%) | Biology (3.9%) |

*Table 1: Ranking of the top four university disciplinary areas (UDA) by university-industry collaboration*

| SDS | UDA | Number of articles |
|---|---|---|
| Electronics | Industrial and inf. engineering | 114 |
| Internal medicine | Medicine | 109 |
| Pharmacology | Biology | 94 |
| Biochemistry | Biology | 89 |
| Industrial chemistry | Chemistry | 78 |
| Organic chemistry | Chemistry | 76 |
| Experimental physics | Physics | 74 |
| General and inorganic chemistry | Chemistry | 73 |
| Pharmaceutical chemistry | Chemistry | 69 |
| Physical chemistry | Chemistry | 60 |

*Table 2: Ranking of top-ten scientific disciplinary sectors (SDS) by number of articles in university-industry co-authorship; the UDA for each sector is indicated in parentheses*

| SDS | UDA | Academic articles in co-authorship with industry, out of total SDS articles (%) |
|---|---|---|
| Energy and environmental systems | Industrial and inf. engineering | 15.0 |
| Polymer materials science and technology | Chemistry | 13.3 |
| Electronics | Industrial and inf. engineering | 12.6 |
| Aerospace installations and systems | Industrial and inf. engineering | 11.1 |
| Primary materials engineering | Industrial and inf. engineering | 11.1 |
| Industrial chemistry | Chemistry | 10.9 |
| Electrical systems for energy | Industrial and inf. engineering | 10.9 |
| Hydrocarbons and ground fluids | Industrial and inf. engineering | 10.0 |
| Applied physical chemistry | Industrial and inf. engineering | 9.8 |
| Electrical and electronic measurement | Industrial and inf. engineering | 9.7 |

*Table 3: Top-ten SDS ranking by percentage of university-industry co-authored articles out of total SDS articles*

| SDS | UDA | Academic articles in co-authorship with industry, out of total SDS co-authored articles (%) |
|---|---|---|
| Energy and environmental systems | Industrial and inf. engineering | 30.0 |
| Manufacturing technology and systems | Industrial and inf. engineering | 21.4 |
| Electrical energy systems | Industrial and inf. engineering | 21.2 |
| Commodities engineering | Industrial and inf. engineering | 20.0 |
| Applied physical chemistry | Industrial and inf. engineering | 17.9 |
| Electronics | Industrial and inf. engineering | 17.8 |
| Electrical and electronic measurement | Industrial and inf. engineering | 17.7 |
| Environmental chemistry | Chemistry | 17.1 |
| Aerospace construction and structures | Industrial and inf. engineering | 16.7 |
| Aerospace systems and plants | Industrial and inf. engineering | 16.7 |

*Table 4: Top-ten SDS ranking according to the percentage of university-industry co-authored articles out of total SDS co-authored articles*

| SDS | UDA | Academic articles in co-authorship with industry per SDS researcher |
|---|---|---|
| Polymer materials science and technology | Chemistry | 1 |
| Industrial chemistry | Chemistry | 0.53 |
| Electronics | Industrial and inf. engineering | 0.35 |
| Applied physical chemistry | Industrial and inf. engineering | 0.34 |
| Chemical fundaments of technology | Chemistry | 0.28 |
| Principles of chemical engineering | Industrial and inf. engineering | 0.27 |
| Environmental and cultural property chemistry | Chemistry | 0.19 |
| Materials science and technology | Industrial and inf. engineering | 0.19 |
| Applied pharmaceutical technology | Chemistry | 0.18 |
| Molecular biology | Biology | 0.17 |

*Table 5: Top-ten SDS ranking by academic articles in co-authorship with industry per SDS researcher*

| | $IF_{pr}$ of all publications | $IF_{pr}$ of publications from collaboration |
|---|---|---|
| Average | 56.59 | 58.62 |
| Variance | 80.55 | 88.93 |
| N. observations * | 174 | 174 |
| Stat-t | -2.053 | |
| p-value (one tail) | 0.04 | |
| p-value (two tail) | 0.020 | |

*Table 6: Comparison of impact factor of all publications and for all those realized in collaboration*
* The comparison excludes 7 SDSs with less than 7 publications from collaboration

|  | $IF_{pr}$ of all publications | $IF_{pr}$ of publications in collaboration with private sector |
|---|---|---|
| Average | 57.22 | 56.16 |
| Variance | 71.48 | 273.75 |
| N. observations ** | 141 | 141 |
| Stat-t |  | 0.673 |
| p-value (one tail) |  | 0.501 |
| p-value (two tail) |  | 0.250 |

*Table 7: Comparison of impact factor of all publications and for those realized in collaboration with the private sector*
** The comparison excludes SDSs where there was no publication in collaboration with the private sector.

| | $O_{pr}$ | | $FSC_{pr}$ | |
|---|---|---|---|---|
| *Statistics* | Scientists who collaborated at least once with private sector researchers | Rest of the population | Scientists who collaborated at least once with private sector researchers | Rest of the population |
| Average | 71.94 | 56.72 | 64.79 | 50.08 |
| Variance | 540.76 | 632.57 | 759.01 | 828.46 |
| N. observations | 1,705 | 16,152 | 1,705 | 16,152 |
| Stat-t | 28.211 | | 23.107 | |
| p-value (one tail) | 2 E-149 | | 1 E-105 | |
| p-value (two tail) | 2 E-149 | | 1 E-105 | |

*Table 8: Comparison between researchers who collaborated with industry at least once (2000-2003 triennium) and the rest of the population*

| | $Ii\_SDS$ | | $Ii\_SCI$ | |
|---|---|---|---|---|
| *Statistics* | All publications | Those in co-authorship with industry | All publications | Those in co-authorship with industry |
| Average | 1.649 | 1.763 | 2.111 | 2.222 |
| Variance | 0.070 | 0.366 | 0.260 | 0.507 |
| N. observations | 141* | | 144** | |
| Stat-t | -2.063 | | -1.519 | |
| p-value (one tail) | 0.020 | | 0.065 | |
| p-value (two tail) | 0.040 | | 0.130 | |

*Table 9: Indexes of multidisciplinarity for publications by Italian university and those in collaboration with private companies*
* The comparison excludes SDSs where there was no publication in collaboration with the private sector.
** The comparison excludes ISI's categories where there was no publication in collaboration with the private sector.

| Statistics | $Ii_{\_SDS}$ | | $Ii_{\_SCI}$ | |
| --- | --- | --- | --- | --- |
| | All publications in collaboration | Those in co-authorship with industry | All publications in collaboration | Those in co-authorship with industry |
| Average | 1.696 | 1.763 | 2.114 | 2.222 |
| Variance | 0.065 | 0.366 | 0.276 | 0.507 |
| N. observations | 141 | | 144 | |
| Stat-t | -1.219 | | -1.463 | |
| p-value (one tail) | 0.112 | | 0.072 | |
| p-value (two tail) | 0.224 | | 0.145 | |

*Table 10: Indexes of multidisciplinarity for academic publications in collaboration and those in co-authorship with private companies*